\documentclass{article}
\usepackage{graphics}
\begin{document}

%% 1. Existing Editorial Commands
\newcommand{\no}{\noindent}
\newcommand{\hf}{\hfill}
\newcommand{\be}{\begin{equation}}
\newcommand{\ee}{\end{equation}}
\newcommand{\bc}{\begin{center}}
\newcommand{\ec}{\end{center}}
\newcommand{\bea}{\begin{eqnarray}}
\newcommand{\eea}{\end{eqnarray}}
\newcommand{\non}{\nonumber}
\newcommand{\fns}{\footnotesize\ }
\newcommand{\nms}{\normalsize\ }
\newcommand{\mult}{\multicolumn}

%% 2. Made-up Editorial commands
\newcommand{\vs}{\vspace{2.5mm}}
\newcommand{\vn}{\vs\no}
\newcommand{\hs}{\hspace{2.5mm}}
\newcommand{\para}[1]{\vs\vs\bc{\bf#1}\ec}
\newcommand{\rf}[1]{$^{[#1]}$}   

%% 3. Existing Symbols
\newcommand{\rarr}{\rightarrow}
\newcommand{\larr}{\leftarrow}
\newcommand{\pll}{\parallel}  %% \perp is already in Latex
\newcommand{\p}{\partial}

%% 3.1 Existing Symbols Made Amphibious
\newcommand{\x} {\mbox{$\times$}}
\newcommand{\up}[1]{\mbox{$^{#1}$}}
\newcommand{\dn}[1]{\mbox{$_{#1}$}}
\newcommand{\bm}[1]{\mbox{\boldmath $#1$}} 

%% 4. Made-up symbols 
\newcommand{\abs}[1]{\left| #1 \right|} 
\newcommand{\siml}{\stackrel{<}{\sim}} 
\newcommand{\simg}{\stackrel{>}{\sim}} 
\newcommand{\dg}{\mbox{$^\circ$}}
\newcommand{\sun}{\mbox{$_\odot$}}
\newcommand{\ti}[1]{\mbox{$\times 10^{#1}$}}
\newcommand{\av}[1]{\mbox{$\langle{#1}\rangle$}}
\newcommand{\pp}[2]{\mbox{${\frac{\p#1}{\p#2}}$}}
\newcommand{\uph}{\up{h}}
\newcommand{\upm}{\up{m}}
\newcommand{\ups}{\up{s}}
\newcommand{\hms}[3]{\mbox{#1\uph #2\upm #3\ups}}
\newcommand{\dms}[3]{\mbox{#1\dg #2$'$ #3$''$}}

%%  5. Romans 
\newcommand{\rmd}{{\rm d}}
\newcommand{\eff}{\mbox{\rm eff}}
\newcommand{\rme}{{\rm e}}
\newcommand{\mod}{{\rm mod}}
\newcommand{\au}{{\sc au}}
\newcommand{\alfv}{Alfv\'{e}n}

%%  6. Greeks
\newcommand{\alf}{\mbox{$\alpha$}}
\newcommand{\bet}{\mbox{$\beta$}}
\newcommand{\gam}{\mbox{$\gamma$}}
\newcommand{\dlt}{\mbox{$\delta$}}
\newcommand{\Dlt}{\mbox{$\Delta$}}
\newcommand{\Nbl}{\mbox{$\nabla$}}
\newcommand{\eps}{\mbox{$\epsilon$}}
\newcommand{\veps}{\mbox{$\varepsilon$}}
\newcommand{\vphi}{\mbox{$\varphi$}}
\newcommand{\lam}{\mbox{$\lambda$}}
\newcommand{\muu}{\mbox{$\mu$}}
\newcommand{\mum}{\mbox{$\mu$m}}
\newcommand{\mus}{\mbox{$\mu$s}}
\newcommand{\nuu}{\mbox{$\nu$}}
\newcommand{\sgm}{\mbox{$\sigma$}}
\newcommand{\ha}{H\dn{\alf}}
\newcommand{\hb}{H\dn{\bet}}
\newcommand{\hg}{H\dn{\gam}}
\newcommand{\hd}{H\dn{\dlt}}

%% 7. Complex Units
\newcommand{\ksm}{\mbox{km\,s\up{-1}\,Mpc\up{-1}}}
\newcommand{\gcm}{\mbox{g\,cm\up{-3}}}
\newcommand{\jcs}{J\,cm\up{-2}\,s\up{-1}}
\newcommand{\esh}{erg\,s\up{-1}\,Hz\up{-1}}
\newcommand{\ecsh}{erg\,cm\up{-2}\,s\up{-1}\,Hz\up{-1}}

\newcommand{\nfc}{\mbox{$N_{\rm fc}$}}
\newcommand{\nec}{\mbox{$N_{\rm ec}$}}
\newcommand{\dxo}{\mbox{$\delta x_0$}}
\newcommand{\dyo}{\mbox{$\delta y_0$}}
\newcommand{\dzo}{\mbox{$\delta z_0$}}

\no{\Large\bf Rhombic Cell Analysis. II. Application to the IRAS/PSCZ
Catalogue}

\vn\vn{\large T. Kiang\up{1}\hs\hs\hs Y.F. Wu\up{2}\hs\hs\hs
X. F. Zhu\up{2}}

\vn \up{1} Dunsink Observatory, Dublin Institute for Advanced Studies,
Dublin 15, Ireland

\vn \up{2} Center for Astrophysics, University of Science and
Technology of China, Hefei, 230026

\begin{quotation}
{\bf Abstract}~~Rhombic cell analysis as outlined in the
first paper of the present series is applied to samples of
varying depths and liming luminosities of the IRAS/PSCz
Catalogue. Numerical indices are introduced to summarize
essential information. Because of the discrete nature of the
analysis and of the space distribution of galaxies, the
indices for a given sample must be regarded as each having an
irreducible scatter. Despite the scatter, the mean indices
show remarkable variations across the samples. The underlying
factor for the variations is shown to be the limiting
luminosity rather than the sampling depth. As samples of more
and more luminous galaxies are considered over a range of
some 2.5 magnitudes (a factor of some 75 in space density), the
morphology of the filled and empty regions 
defined by the galaxies 
degrades steadily towards insignificance, and the degrading
is faster for the filled than the empty region.
\end{quotation}

\vn{\bf 1~~~INTRODUCTION}

\vs This is the second paper in the series ``Rhombic Cell
Analysis''.  Here we apply the technique described in Paper I
(Kiang 2003) to the IRAS/PSCz Catalogue (Saunders et al.\
2000, `PSCz' hereinafter). The wealth of data in PSCz has
meant that we can now apply the analysis to various depths
(Section 2), and simultaneously to various limiting
luminosities, since PSCz is flux-limited. The method
developed so far centers on two number distributions, one of
$n_1$, the number of {\it like\/} neighboring cells to a
given cell, and one of $\tau$, a two-suffixed topological
type, the two suffixes being the numbers of {\it like\/} and
{\it unlike neighbor- groups\/}. Here in Section 3 we
introduce four numerical indices, $\eta$ on one hand, and
$\chi_1$, $\chi_2$, $\chi_{21}$ on the other, intended to
respectively summarise the most important characteristics of
the two distributions.  In Section 4 we shall point out an
important circumstance in the practical application, namely,
the $n_1$- and $\tau$- distributions depend rather sensitively
on the precise location of the zero of the cells and that, as
a consequence, each of the indices introduced above must be
regarded as a random variable with a certain probability
distribution. For each chosen sample of galaxies we consider
a set of 16 independent zero offsets of the analysing cells
and calculate the mean and standard deviation 
of the four indices.  And we
do this separately for the sets of filled cells and empty
cells.  In Section 5 we plot the mean indices as functions of
the sample depth, $r_{\ast}$. Remarkable features are found
in the curves and their interpretation  concludes this paper.

\vspace{5mm}

\no{\bf 2~~~THE IRAS/PSCz CATALOGUE}

\vs The IRAS/PSCz Catalogue (Saunders et al.\ 2000) is the most
comprehensive redshift catalogue to date. It covers 84\% of
the sky and contains about 15000 galaxies to a uniform
$\mu_{60}$ flux limit of 0.60 Jy. We counted 14669
galaxies with measured (positive) redshifts $z$.

The 16\% of the sky not covered by PSCz (the ``gaps'')
consists of a large irregular zone of avoidance along the
galactic equator, and two narrow stripes extending to high
latitudes (Saunders et al., Fig.\,4). For our purpose the
gaps must first be filled with mock galaxies. We are most
grateful to Dr Fabio Fontanot of Trieste for kindly providing
us, before publication of the paper where it was used
(Fontanot et al.\ 2003), with just such a list of 2808 mock
galaxies, constructed according to a procedure given in
(Branchini et al.\ 1999).  The Fontanot list gives only the
galactic coordinates and the redshifts; but we shall also be
needing flux values. As we only need a random selection from
the distribution of fluxes in PSCz, and as the Fontanot list
is ordered quite independently of PSCz, we simply assigned
every fifth value in PSCz to the Fontonot list. This
``Filled-out PSCz catalogue'' of $14669+2808=17477$ galaxies
is the basic material for the present study.

We should point out that while a minor part of the data in
the Fontanot list pertaining to the two narrow stripes were
used in our work with the full weight (on an equal footing
with the real PSCz data), the major part pertaining to the
galactic zone of avoidance were used only in a diminished
capacity (See Section 2.3). Similarly, the same diminished
role is played by the small part of the PSCz data with
$|b|<10\dg$.

\vn\vn{\bf 2.1~~Nominal Distance and Luminosity}

We now have an all-sky catalogue of 17477
galaxies, each with a flux value and a redshift value. The
next thing we do is to convert the redshifts into distances.
For our purpose we can simply use a ``nominal'' distance,
given by
%%%%%%%%%(1)
\be  r=cz/100 (r\;\mbox{in Mpc}, cz\;\mbox{in km/s})  \ee
%%%%%%%%%
\no This is because the subject-matter of our analysis, the
number distributions of $n_1$ and of $\tau$, ultimately depend
only on the numbers of galaxies located inside a given
(rhombic) cell and inside its twelve neighboring cells, so if
we adopt a certain cosmology and use sophisticated formulae
(of comoving distance for the cells, and of luminosity
distance for the galaxies), the effect of so doing on the
$n_1$- and $\tau$-distributions can be expected to be quite
small, considering that PSCz only extends to $z\sim 0.1$.

With $r$ so defined, we evaluate, for each galaxy, its usual
rectangular coordinates $x=r\cos b\cos l, y=r\cos b\sin l,
z=r\sin b$, and then, together with its flux $f$, its
(nominal) luminosity $L$ defined simply as $L=f\cdot r^2$
(other authors may prefer to add a factor of 4$\pi$ here), or
%%%%%%%%%(2)
\be 
    \log L = \log f + 2\log r  \;,  
\ee
\no where $f$ is in Jy, and $r$ is in Mpc.

\vn\vn{\bf 2.2~~Distance {\it and\/} luminosity limited Samples}

In Paper I, the analysis was applied to the CfA catalogue,
regarded as a single sample. At the end of that paper, the
hope was expressed that with larger survey results becoming
available, much finer analysis can be made. The PSCz has
provided just such an opportunity.

PSCz is flux-limited. This means that within a
distance-limited sample taken from PSCz, the coverage in
luminosity is not uniform: it decreases with increasing
distance. For uniformity, then, we should not take {\it
all\/} the galaxies up to a certain distance $r_{\ast}$, but only
those with luminosities greater than the corresponding
limiting luminosity $L_{\ast}$, given by
%%%%%%%%%(3)
\be
     \log L_{\ast} = \log 0.60  + 2 \log r_{\ast}\;.
\ee
\no Thus, each basic unit sample of our analysis satisfies
{\it two\/} conditions, $r<r_{\ast}$ and $L>L_{\ast}$. In the $\log L
\sim 2\log r$ plane, such a sample is represented by a
rectangular box with its lower right corner on the limiting
diagonal. Such boxes generally overlap.

Our analysis is applied to 8 such samples (Table 1).  Each
sample is identified by its limiting distance $r_{\ast}$, e.g.,
S-75, S-150, S-250 (Column 1).  Their defining limiting
luminosities are listed in Column 2.  The eight samples cover
a range of 0.52 dex in $r_{\ast}$, and a range of 1.04 dex (or 2.6
magnitudes) in $L_{\ast}$.

 \bc
 {\small\bf {Table 1} Eight Distance- {\it and\/} Luminosity-
 Limited Samples}\\
 \fns
 \begin{tabular}{cc|cccc}

Sample & $\log L_{\ast}$ & $a_0$ & \nfc=\nec &
$a_{0,\min}$---$a_{0,\max}$ & $\rho$ (10\up{-6}\,Mpc\up{-3})  \\
 (1)   &   (2)  &   (3)   &  (4)  &     (5)        & (6) \\ \hline\hline

S- 75  &  3.53  &  7.173  &  780  &  7.072-- 7.296  & 1441.0 \\
S-100  &  3.78  &  9.140  &  948  &  8.970-- 9.140  &  631.0 \\
S-125  &  3.97  & 11.137  & 1020  & 11.034--11.204  &  301.0 \\
S-150  &  4.13  & 13.856  &  925  & 13.730--14.220  &  150.0 \\
S-175  &  4.26  & 16.762  &  797  & 16.393--16.762  &   84.9 \\
S-200  &  4.38  & 19.750  &  737  & 19.340--19.982  &   49.3 \\
S-225  &  4.48  & 22.970  &  660  & 22.898--23.340  &   29.4 \\
S-250  &  4.57  & 27.620  &  492  & 27.070--27.620  &   18.3 \\

 \end{tabular}
 \normalsize
 \ec

\vn\vn{\bf 2.3 An Illustrative Example}

We take S-150 to illustrate the various steps in our
calculation.  First, to improve homogeneity of data, we
define ``actual used cells'' by imposing, beside the
condition $r<150$, two further restrictions on the
coordinates of the cell centres. 1) To avoid the large
uncertainties in the inferred distances of the galaxies in our
local ``swimming pool'', we require $r>25$. 2) To reduce
undue influence of the mock galaxies in the galactic zone of
avoidance, we require all the cells should be completely
above latitude 10\dg. Now, our cells which can each be
imagined as consisting of a cube of sides $a_0$ with its six
faces covered by six pyramids of height $a_0/2$, are arranged
along the galactic coordinate axes, so the second condition
reads,
%%%%%%%(4)
\be
 |z| > z_{\rm lim} + a_0,\;   {\rm where}\;
                         z_{\rm lim}=r\,\sin 10\dg\,.
\ee
Thus, of the mock galaxies  below 10\dg\ galactic latitudes,
we made use of only those located in cells that share a
common face with some cells completely inside the boundary,
and then only in so far as they contribute to the definition
of the like/unlike status of the common face.  This remark
applies also to those PSCz galaxies below latitude 10\dg.

With respect to the ``outer'' boundary of $r=150$ in the
present example of S-150, the need is obviously not so
compelling that the actual used cells should be {\it
completely\/} inside that boundary;---to require this would
mean a substantial drop in the number of usable cells.  So we
define usable cells as those whose {\it centres\/} are within
the $r=150$ limit.  Then our usable cells on the boundary
will have parts outside that boundary, and moreover, the
status of the ``boundary'' faces will further depend on the
presence or absence of galaxies inside cells in the next
shell out. It is easy to show that, for a given $a_0$, if we
consider all galaxies to a distance of 
$150 + (\sqrt{2}+1)a_0$, 
then we will have included all relevant galaxies.

We find, by trial and error, the value of $a_0$ that will
give equal numbers (or as equal as possible) of filled cells
and empty cells, \nfc\ and \nec\ of the actual usable cells.
It is understandable that the two numbers are highly
discontinuous functions of $a_0$, and often a coarse
adjustment succeeds where a fine-tuning fails. In the present
case we found that $a_0=13.856$ succeeded in giving
$\nfc=\nec=925$ (Table 1, Cols. 3,4).

We can now summarise the various steps in this particular
example of S-150.  We start with the `filled-out PSCz
catalogue' of 17477 galaxies, and pick out those with $\log L
\geq 4.13$ (Table 1, Line 4, Column 2). Then, for each trial
value of $a_0$, we determine the filled/empty status of all
the cells with centres out to distance $150 + \sqrt{2} a_0$.
Then, for the ``actual used cells'' with centres satisfying
the three conditions, $25\leq r$, $r \leq 150$, and the
inequality (4), we count the number of filled cells, \nfc,
and the number of empty cells, \nec.  We vary $a_0$ until
\nfc\ and \nec\ are equal (or as nearly equal as possible).
Then for this optimal value of $a_0$, we finally obtain the
object of our analysis, the separate $n_1$- and
$\tau$-distributions for the filled and empty cells.

The actual results in the present example are displayed in
Figs.\,1 and 2. They can be compared with the Figures 1 and 2
of Paper I. The final results of the present paper will be based on
$8\x 16=128$ such pairs of figures.

%%%%%%  Figures 1 and 2  to be approximately
%%%%%%  here.  Best made half-with and shown side-by-side
%%%%%%  like the figs. 1 and 2 of Paper I, ChJaa 3/2 p.102

\begin{figure}
  \centering {\includegraphics{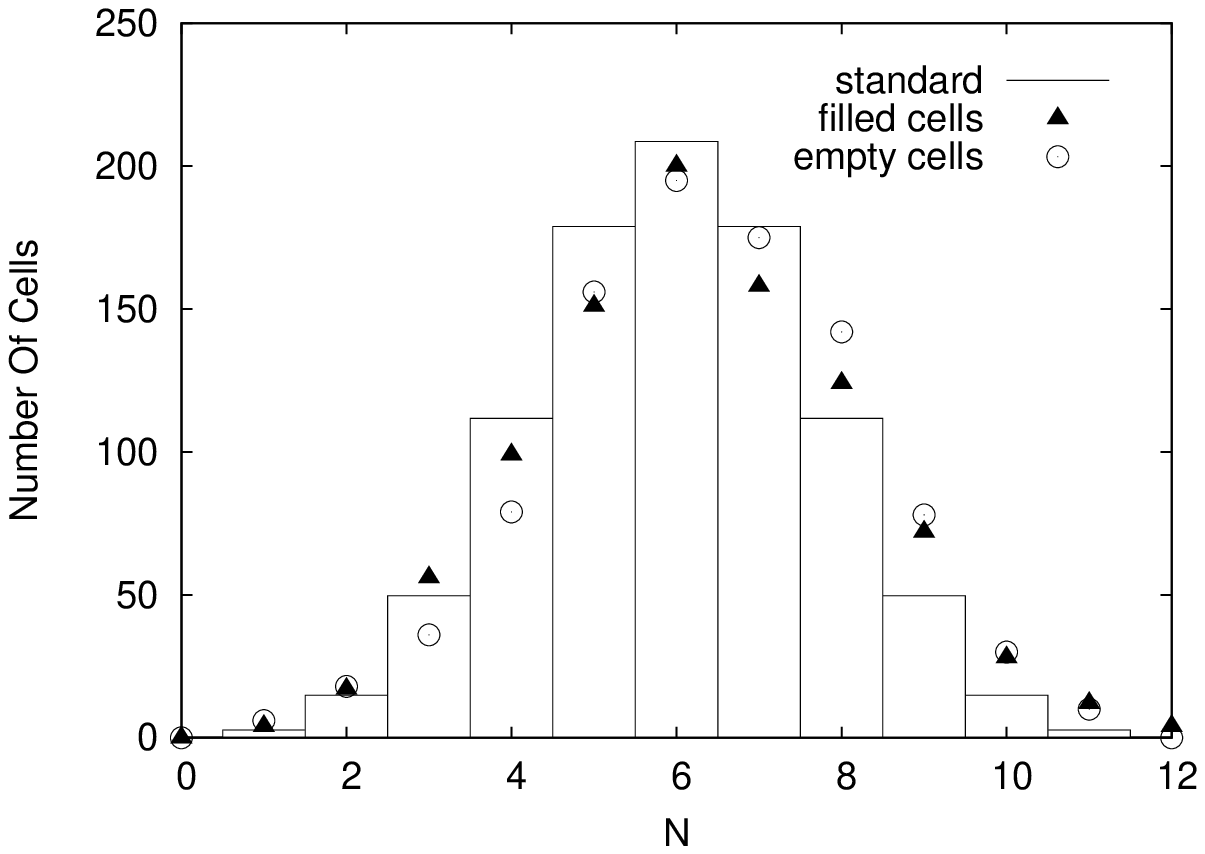}}
  \bc{\small Fig.\,1~~The $n_1$-distriubtion of the
 Illustrative Example. Histogram is the binomial
distribution for the case of pure random mixture
of filled and empty cells}\ec
\end{figure}

\begin{figure}
  \centering {\includegraphics{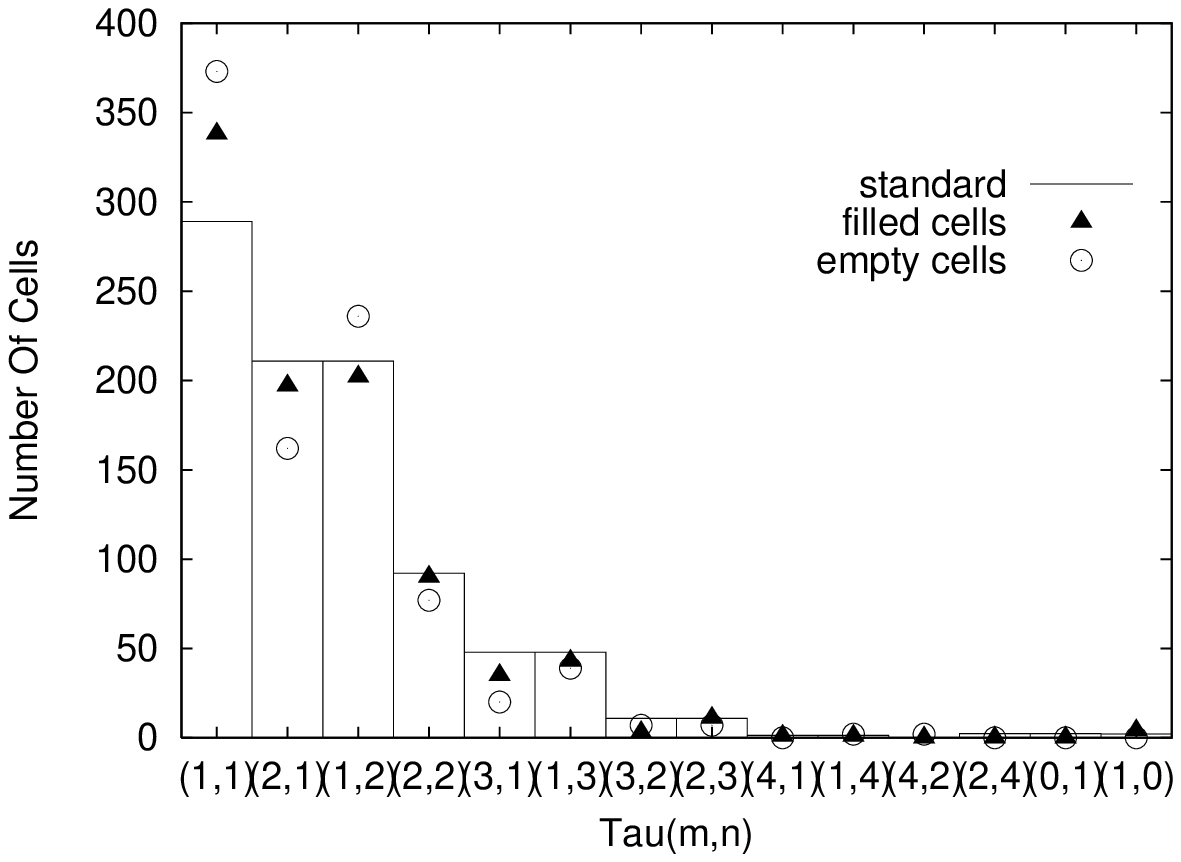}}
  \bc{\small Fig.\,2~~The $\tau$-distribution of the
Illustrative Example. Histogram corresponds to the case of
pure random mixture of filled and empty cells}\ec
\end{figure}

The values of the optimal $a_0$ for all the eight samples are given
in Column 3 of Table 1. It happened that in all cases we had
\nfc$=$\nec. These equal values are listed in Column 4.

\vspace{5mm}

\no{\bf 3~~~STATISTICAL INDICES}

\vs Our aim being comparison between the filled and empty regions
of the universe based on their $n_1$- and $\tau$-
distributions, it is obviously advantageous if we can
summarise the most salient features of the distributions by
some numerical indices.  We have singled out one index, 
dubbed 
the ``flocking index'', for the $n_1$- distribution and
three ``$\chi$-type'' indices for the $\tau$-distribution.

\vn\vn{\bf 3.1~~The Flocking Index $\eta$}

It was pointed out in Paper I (Section 3.2.1) that, because
of the imposed condition \nfc=\nec, we must use some function
of the observed $n_1$ frequencies other than the mean value
$<n_1>$ as the ``flocking index'' $\eta$ that quantifies the
degree to which cells of the same kind (filled or empty)
flock together.

Let us first rationalize the $n_1$- distribution as follows:
(i) We re-center the distribution at 0, i.e., we use a new
independent variable, $k=n_1-6$, ($k=0,\pm 1,\pm 2\cdots,\pm
6$). (ii) Because the frequency in the two end boxes ($k=\pm
6$) is usually very small, we put it into the next box, and
regard the combined frequency as located at the ``binomial''
mean of the two, so now we have the rationalised distribution
$N(k)$, defined for $k=0,\pm 1,\pm 2,\pm 3,\pm 4,\pm 5.228$.

Now, consider the five differences, $ d_k\equiv
N(|k|)-N(-|k|), |k|=1,2,3,4,$ 5.228. Naturally, we would want our
$\eta$ to be positive when the $d_k$ are generally more
positive than negative; however, a straight sum of the five
$d_k$ would not be proper:  they should each first be {\it
standardized\/} by their expected random error. It seems
reasonable, as a first approximation, to regard $N(|k|)$ and
$N(-|k|)$ as two independent Poisson variables; then the
variance of their difference would just be equal to their
sum. Hence we standardize each difference by the square root
of their sum, thus, $D(k) \equiv (N(|k|)-N(-|k|))/
\sqrt{N(|k|)+N(-|k|)}$.
But a straight sum of the $D(k)$
still does not seem quite right. Recalling the
meaning of $k$, it seems reasonable that these standardized
differences should each be given a weight equal to $k$. Thus,
we finally arrive at our adopted formula for the flocking
index $\eta$:
%%%%%%(5)
\be
   \eta = \sum k D(k) / \sum k\,,  \; k=1,2,3,4,5.228 \,.
\ee
\no In words, $\eta$ is a weighted
average of standardized difference
between the observed frequences at equal and opposite
distances from the centre of the $n_1$- distribution,
standardized with the expected random error of the
difference, and weighted according to the distance from the
centre. Note, the observed central frequency at $k=0$ (or 
$n_1=6$) does not enter into the calculation of ${\eta}$.

\vn\vn{\bf 3.2~~The Three $\chi$- type Indices}

Our $\tau$- distribution, being the number distribution of a
two-parametered variable $\tau(m_1, m_2)$, does
not lend itself to be similarly summarised by a single
statistic. However, as was pointed out in Paper I, at the
present stage of development of the rhombic cell
analysis, we should perhaps concentrate on just 2 of the entries,
namely, the two pertaining to $\tau(2,1)$ and $\tau(1,2)$,
for these are respective signatures of one-ply strings and
monolayers (Paper I, 3.2.2). Writing for short, the
observed frequency of $\tau(2,1)$ as $N(2,1)$, and its
expected frequency (expected on the assumption of a
thorough mixture of filled and empty cells) as $E(2,1)$, we
define statistic
%%%%%(6)
\be
   \chi_1 = (N(2,1)-E(2,1))/ \sqrt{E(2,1)}\,,
\ee
\no as a measure of the degree to which the observed
frequency of one-ply string cells exceeds its random
expectation.

Similarly, we define statistic
%%%%%%(7)
\be
   \chi_2 = (N(1,2)-E(1,2))/ \sqrt{E(1,2)}\,,
\ee
\no in regard to the observed frequency of monolayer cells.

We have found it useful to introduce a third statistic that
quantities the difference $N(1,2)-N(2,1)$. Regarding the two
as independent random variables, the proper standardized
difference is
%%%%%%(8)
\be
    \chi_{21} = (N(1,2)-N(2,1)) / \sqrt{N(1,2)+N(2,1)}\,,
\ee
\no which can be taken as a measure of the excess of ``sheet
cells'' over ``string cells''.

\vspace{5mm}

\no{\bf 4~~~ZERO OFFSETS OF THE GRID OF CELLS}

\vs The present, more detailed application of rhombic cell
analysis revealed a most important fact, which had no
occasion of emerging in the preliminary application reported
in Paper I.  Namely, the $n_1$- and $\tau$- distributions are
highly sensitive to the exact placing of the grid of rhombic
cells, that is, to the zero offset.  To explain, recall
that our rhombic cells are defined in the following manner:
space is first partitioned into a three-dimensional
chessboard of black and white cubes of sides $a_0$, all the
white cubes are each cut into six pyramids, and a rhombic
cell consists of a black cube with six white pyramids stuck
on its faces. The centre of the rhombic cell coincides with
the centre of the black cube. We label the cells by integer
triplets ($i,j,k$), with $i,j,k=0, \pm 1, \pm 2, \cdots$,
subject to $i+j+k=0$ mod(2). We imagine the cells to form a
rigid 3-d frame.  We start our calculations by placing the
centre of our zeroth cell (0,0,0) at galactic coordinates
(0,0,0). What we found was that if we displace our entire
grid of cells by an amount up to and including one
unit of $a_0$, in any combination of the three directions,
then the resulting $n_1$- and $\tau$- distributions will
generally be different, sometimes greatly so.

Now, the maximum displacement is one unit of $a_0$
either along one of the three axes, or along all three axes,
(but NOT along two of the axes, which would reproduce the
original grid). We denote this maximum displacement
by (1,1,1), and the original grid by (0,0,0).

Let us now consider displacements involving half-units of
$a_0$, either positive or negative, in either 1, 2, or 3 of
the 3 directions. There are altogether $3\x 3\x 3 -1= 26$
such displacements or offsets. But not all 26 are
independent: some reproduce the same displacement of the
grid.  More precisely, the same displacement is obtained if
we reverse the signs of any two non-zero displacements,
e.g., the displacements or offsets (+,0,-) and (-,0,+) are
duplicates of each other, and so are (+,+,-) and (-,+,+), and
so on. In sum, 12 out of the 26 are duplicates, and 14 are
independents.
%%%%%%%%%
\footnote{The 14 independent offsets can be taken as (0,0,+),
(0,0,-), (0,+,0), (0,-,0), (+,0,0), (-,0,0), (0,+,+),
(0,+,-), (+,0,+), (+,0,-), (+,+,0), (+,-,0), (+,+,+),
(-,-,-)}
%%%%%%%%%
Adding the original grid (0,0,0) and the single one
whole-unit displacement (1,1,1), we have a total of 16
independent displacements or zero offsets.

Of course, so far we have been considering only displacements that
involve half-units of $a_0$. If, for example, we consider
displacements involving quarter-units of $a_0$, the number of
indepenedent displacements will then be much greater.

\bc {\small{\bf Table 2}~~Smallest and Largest Values Among
the 16 Independent Offsets}\\

\fns
\begin{tabular}{ccc|cc|cc}

Sample & $\log L_{\ast}$ & f/e & \multicolumn{2}{c}{$\eta$}
 & \multicolumn{2}{c}{$\chi_{21}$} \\
 &  &  & min & max & min & max  \\ \hline\hline

S- 100  & 3.78 & f & 2.12 & 3.63  & 3.14 & 7.40 \\
        &      & e & 2.81 & 4.23  & 2.80 & 5.80 \\  \hline
S- 200  & 4.38 & f &-1.17 & 0.54  &-2.95 & 2.50 \\
        &      & e &-0.37 & 1.56  &-0.92 & 2.60 \\  \hline

\end{tabular}
\normalsize
\ec

Each displacement or offset calls for a new evaluation of the
optimal $a_0$. We list, in Column (5) of Table 1, the largest
and smallest $a_0$ among the 16 offsets for each of the
considered samples.  Each fresh value of $a_0$ results in
fresh $n_1$- and $\tau$- distributions, and hence in fresh
values of the indices $\eta, \chi_1, \chi_2, \chi_{21}$.
Just to illustrate the effect of the offsets, we display in
Table 2, the smallest and largest values of $\eta$ and
$\chi_{21}$ found among the 16 offsets, for the two samples
S-100 and S-200, and for the filled and empty cells on
separate lines marked ``f'' and ``e''. We note that, in each
case, the range is quite large.

\vn\vn{\bf 4.1~~Irreducible Scatter}

On reflection, we should not be surprised by the fact that
displacing the grid of cells slightly may result in
drastically different $n_1$- and $\tau$- distributions.
Consider, for example, two galaxies that are close together
in space. Then, for one particular placement of the grid, the
two may belong to one and same cell, while slightly shifting
the grid may put them in two different cells; and the $n_1$-
and $\tau$- distributions ultimately depend simply on how
many galaxies go into which cells. We must now recognize the
following fact of life: because of the ultimately
discontinuous nature of the galaxy distribution in space, and
because of the way the rhombic cell analysis works, a given
sample of galaxies does not correspond to some one ``true''
value of an index, rather, it corresponds to a whole
probability distribution of the index. In other words, each
index has an irreducible scatter represented by some
probability distribution, and the 16 values we get from the
16 independent offsets must be regarded as so many
independent random samples taken from that parent
distribution. And the most we can do is to estimate the mean
and standard deviation (s.d.) 
of the parent distributions.  Assuming normal
distribution, the unbiased estimates of the mean and s.d.\
are given by the usual formulae,
%%%%%%%%
\footnote{In a more strict notation, $\av{x}$ would be
written as $\hat{\mu}$, and $\sgm$, as $\hat{\sgm}$}.
%%%%%%%%%
%%%%%(9)
\be \av{x}=\sum x_i /n\,, \;\;
    \sgm=\sqrt{(\sum(x_i-\av{x})^2/(n-1)}\,,\;\;(n=16)\;.
\ee
\no Here, $x$ stands for any one of the four indices and the
summation is over the 16 observed values $x_i$. We emphasize
that, here, $\sgm$ is an unbiased estimate of the s.d. of the
hypothetical parent distribution; it is NOT the s.d. of the
sample mean $\av{x}$, usually known as ``s.e.'' (standard        
error). The latter would be equal to $\sgm/\sqrt{n}$, and so
could be made very small by considering much larger values of
$n$ pertaining to displacements at smaller steps:  its size
thus largely depending on some man-made circumstance, it is
inappropriate as an indicator of some objective scatter. On
the other hand, the $\sgm$ defined in (9), {\it is\/} an
estimate of the irreducible scatter.

Our results, then, consist of the estimated mean and s.d. of
the hypothetical distributions of the four indices, $\eta,
\chi_1, \chi_2, \chi_{21}$, separately for the filled and
empty cells of the eight selected distance/luminosity-limited
samples. The results are listed in Table 3.

\bc {\small{\bf Table 3}~~Results of Calculation for
Selected Samples of PSCz Catalogue}\\
\fns
\begin{tabular}{ccc|cc|cc|cc|cc}

Sample & $\log L_{\ast}$ &f/e
& \multicolumn{2}{c}{$\eta$}
& \multicolumn{2}{c}{$\chi_1$}
& \multicolumn{2}{c}{$\chi_2$}
& \multicolumn{2}{c}{$\chi_{21}$} \\
 & & & mean & s.d. & mean & s.d. & mean & s.d. & mean & s.d.\\\hline\hline

S - 75 & 3.53 & f & 4.21&0.37 &-4.49&0.91 & 3.00&0.78 & 5.46&1.02\\
       &      & e & 4.18&0.46 &-4.11&0.80 & 0.95&0.85 & 3.80&0.78\\  \hline
S- 100 & 3.78 & f & 2.83&0.40 &-3.97&0.72 & 3.34&1.34 & 5.37&1.09\\
       &      & e & 3.51&0.41 &-3.54&0.69 & 2.17&0.81 & 4.26&0.84\\  \hline
S- 125 & 3.97 & f & 1.70&0.48 &-2.62&1.34 & 1.83&0.70 & 2.75&0.96\\
       &      & e & 3.06&0.40 &-2.58&1.11 & 2.18&1.22 & 3.38&1.46\\  \hline
S- 150 & 4.13 & f & 1.39&0.54 &-2.02&0.85 & 1.38&0.83 & 2.29&1.13\\
       &      & e & 1.79&0.48 &-2.21&0.82 & 1.68&0.78 & 2.72&0.79\\  \hline
S- 175 & 4.26 & f & 0.72&0.54 &-0.68&1.15 & 0.32&0.83 & 0.72&1.07\\
       &      & e & 1.78&0.35 &-2.01&0.73 & 1.16&0.92 & 2.24&0.96\\  \hline
S- 200 & 4.38 & f &-0.40&0.55 & 0.29&1.19 &-0.17&1.03 &-0.32&1.34\\
       &      & e & 0.77&0.53 &-0.38&0.77 & 0.46&0.64 & 0.61&0.76\\  \hline
S- 225 & 4.48 & f &-0.73&0.51 & 0.22&0.58 &-0.05&0.94 &-0.51&0.66\\
       &      & e & 1.13&0.58 &-0.64&0.93 & 0.92&1.02 & 1.39&0.80\\  \hline
S- 250 & 4.57 & f &-0.31&0.57 & 0.04&0.90 &-0.08&0.91 &-0.14&1.02\\
       &      & e & 1.08&0.55 &-0.85&1.07 & 0.65&0.84 & 1.12&1.14\\  \hline

\end{tabular}
\normalsize
\ec

\vspace{5mm}

\no {\bf 5~~~RESULTS~~~INTERPRETATION}

\vn\vn{\bf 5.1~~Formal Results}

Remarkable trends emerge when the mean indices of Table
3 are plotted as functions of the sample depth $r_{\ast}$. See
Figures 3, 4, 5. In all the figures, filled triangles refer
to filled cells, and open symbols, to empty cells. The error
bars mark one unit of \sgm, the s.d.\ of the parent
distribution defined at (9). In all cases, a trend
is unmistakable despite the fact that each index for a given
sample has quite a scatter as indicated by the size of the
error bars.

\vn{\it 5.1.1 The $\av{\eta}\sim r_{\ast}$ Curve}

Fig.\,3 shows the mean $\eta$ versus sample depth curve,
separately for the filled cells (filled triangles) and empty
cells (open squares). There are two remarkable features. 1)
Both curves start with definitely positive values at the
smallest sample depth (75 Mpc), then both steadily fall with
increasing depth. By 200 Mpc or so, the ``filled curve'' is
marginally below the zero level of nil flocking, while the
``empty curve'', marginally above. 2) While both curves fall
steadily, the empty curve remains consistently above the
filled curve.  Thus, for both the filled region and the empty
region, there is a definite degree of likes flocking together
at smaller depths, but the flocking decreases to
insignificance around 200 Mpc. And, at any depth, the
tendency of likes flocking together is stronger for the empty
cells than for the  filled cells.

%%%%%   Fig.3  approximately here
\begin{figure}
  \centering {\includegraphics{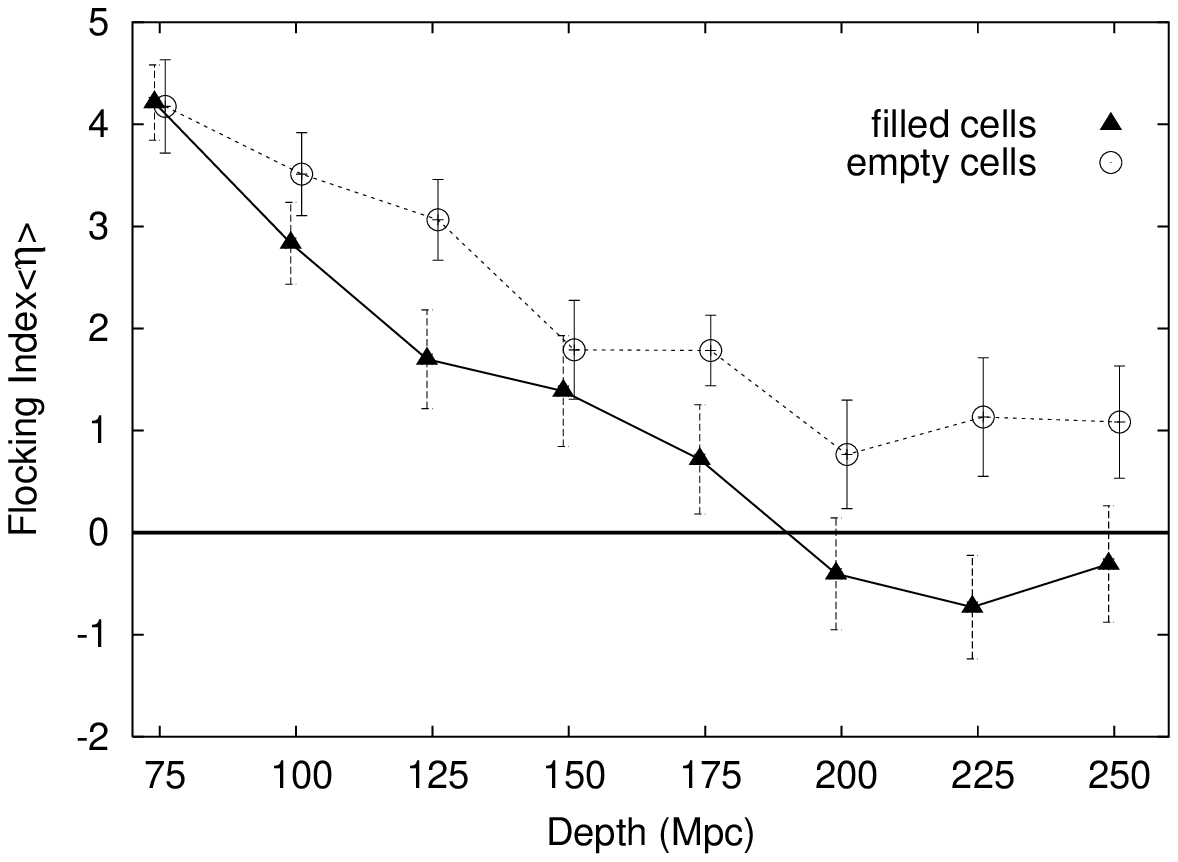}}
  \bc{\small Fig.3~~Mean $\eta$ as a function of the sample depth
for filled and empty cells. Error bars mark 1 s.d. of the
parent distributions}\ec
\end{figure}

Incidentally, for the single (inhomogeneous) CfA sample
studied in Paper I, we have $\av{\eta} =2.09$ for the filled
cells, and 4.02 for the empty cells. These can be seen to
be 
entirely consistent with the curves of Fig.\,1 at depths
$75\sim 100$ and with the last statement.

\vn{\it 5.1.2 The Three $\av{\chi}\sim r_{\ast}$ Curves}

Figs.\,4, 5, 6 show how each of the three $\chi$-indices for
the $\tau$-distribution varies with the sample depth. First,
Fig.\,4 which refers to $\chi_2$, the degree to which the
observed frequency of cells belonging to thin sheets
(strictly, monolayers) exceed the random expectations (Eq.
(7)).

%%%%  Fig.4 approximately here
\begin{figure}
  \centering {\includegraphics{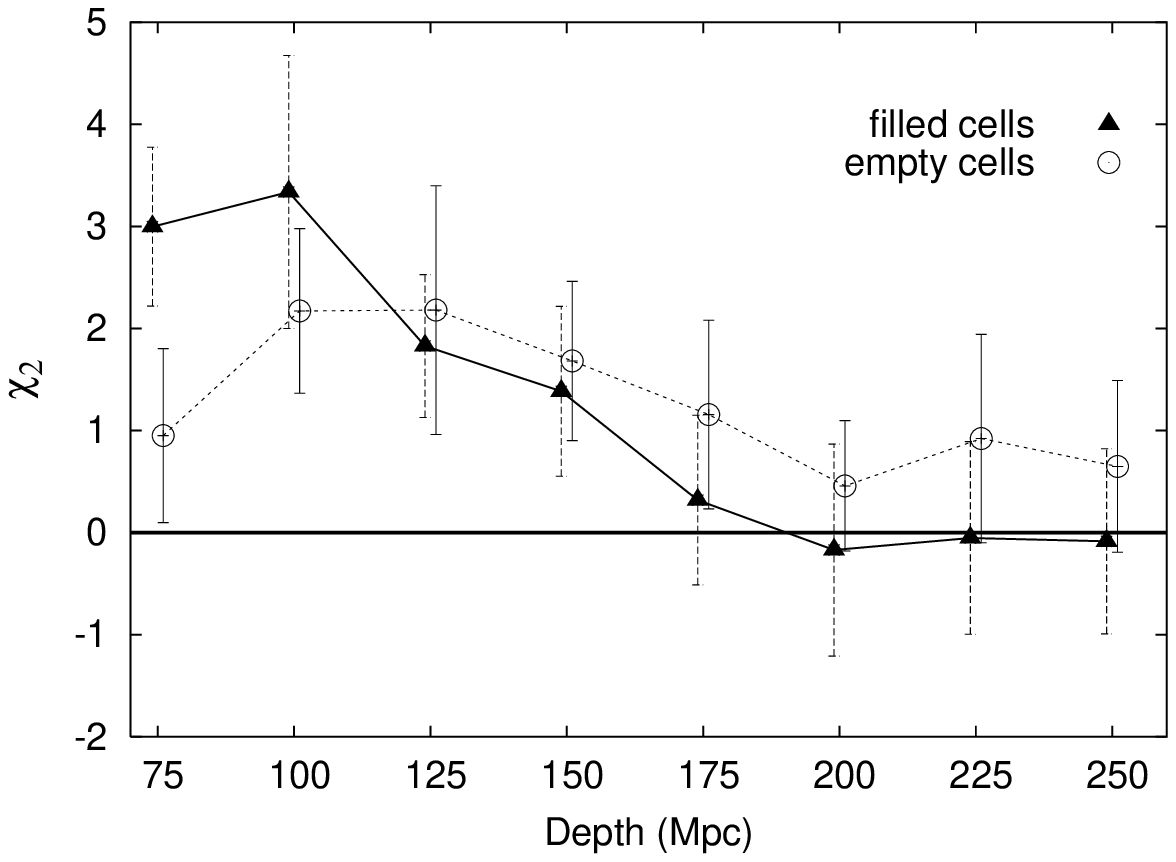}}
  \bc{\small Fig.4~~Mean $\chi_2$ as a function of the sample
depth. Same symbols as in Fig.\,3}\ec
\end{figure}

Fig.\,4 shows: at small depths, both the filled and empty
curves are significantly positive, with the filled lying
definitely above the empty.  This last feature is consistent
with a conclusion reached in Paper I, namely, the filled
cells, but not the empty cells, show a tendency of occurring
in sheets. In fact, we find, for the CfA data,
$\av{\chi_2}=+2.38$ (filled), and -0.92 (empty).  As the sample
depth increases, both curves fall, the filled curve falling
faster, so that at depths around 200 Mpc and beyond, the
filled curve becomes entirely non-significant, while the
empty curve remains marginally significant. The situation at
large depths mimics the behaviour of the $\eta$-curves of
Fig.\,3; note, however, the larger error bars in Fig.\,4.

Next, we consider the mean $\chi_1$ versus depth curves shown in
Fig.\,5.  Recall that $\chi_1$ is a measure of the excess
above random of the observed frequency of cells belonging to
thin (one-ply) strings (Eq.\,(6)).

%%%%%  Fig.5 approximately here
\begin{figure}
  \centering {\includegraphics{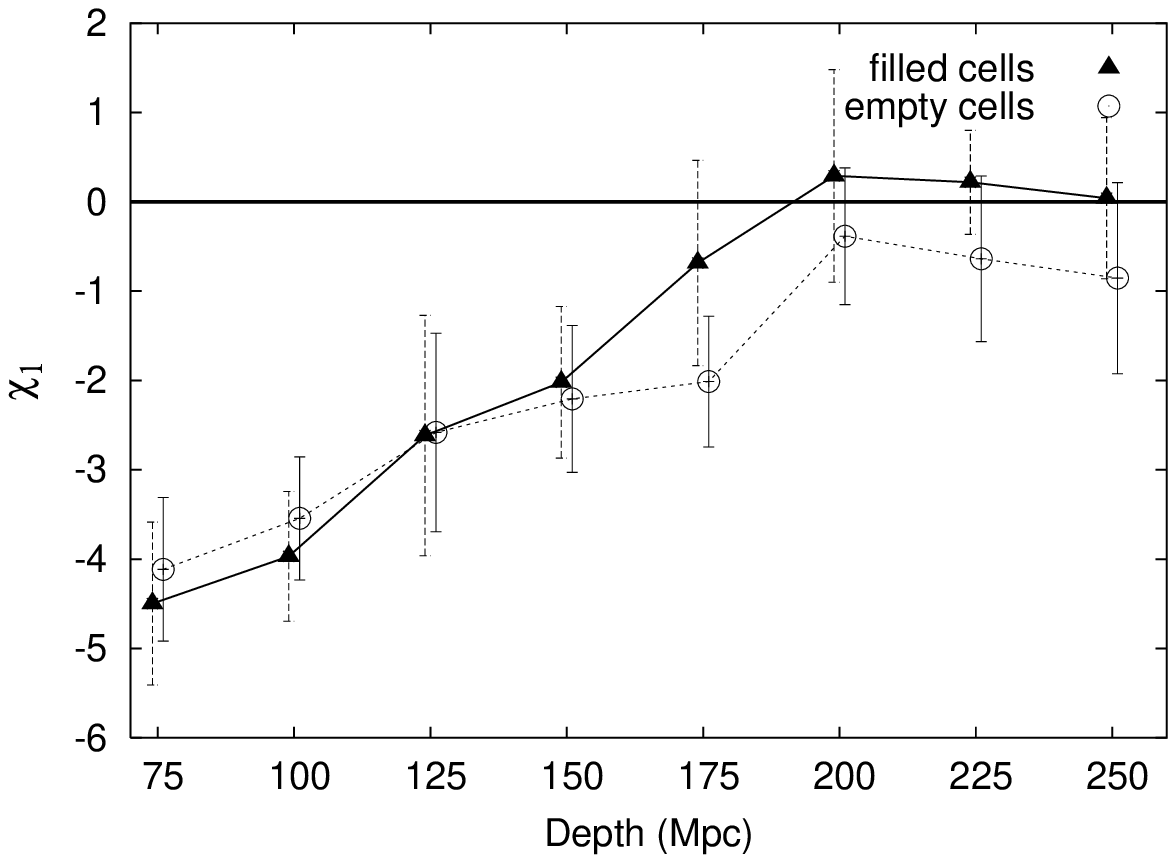}}
  \bc{\small Fig.5~~Mean $\chi_1$ as a function of the sample
depth. Same symbols as in Fig.\,3}\ec
\end{figure}

The $\chi_1$ curves look like some mirror images of the
$\chi_2$ curves: they start significantly below the zero
level and gradually rise to reach it around 200 Mpc, with
the filled curve first below, then eventually above the empty
curve. The values we found for the CfA data are
$\av{\chi_1}=-3.81$ (filled) and -3.28 (empty), entirely
consistent with the initial portions of the curves of
Fig.\,5.

The contrary behaviors of $\chi_2$ and $\chi_1$ suggested to
us that an index for their difference may be of interest.
Hence the index $\chi_{21}$ defined at (8). It could be
called the ``sheet-string differential index'': it measures
the excess of cells belonging to thin sheets over those
belonging to thin strings, with no reference to their common
random expectations.

%%%%%  Fig. 6 approximately here
\begin{figure}
  \centering {\includegraphics{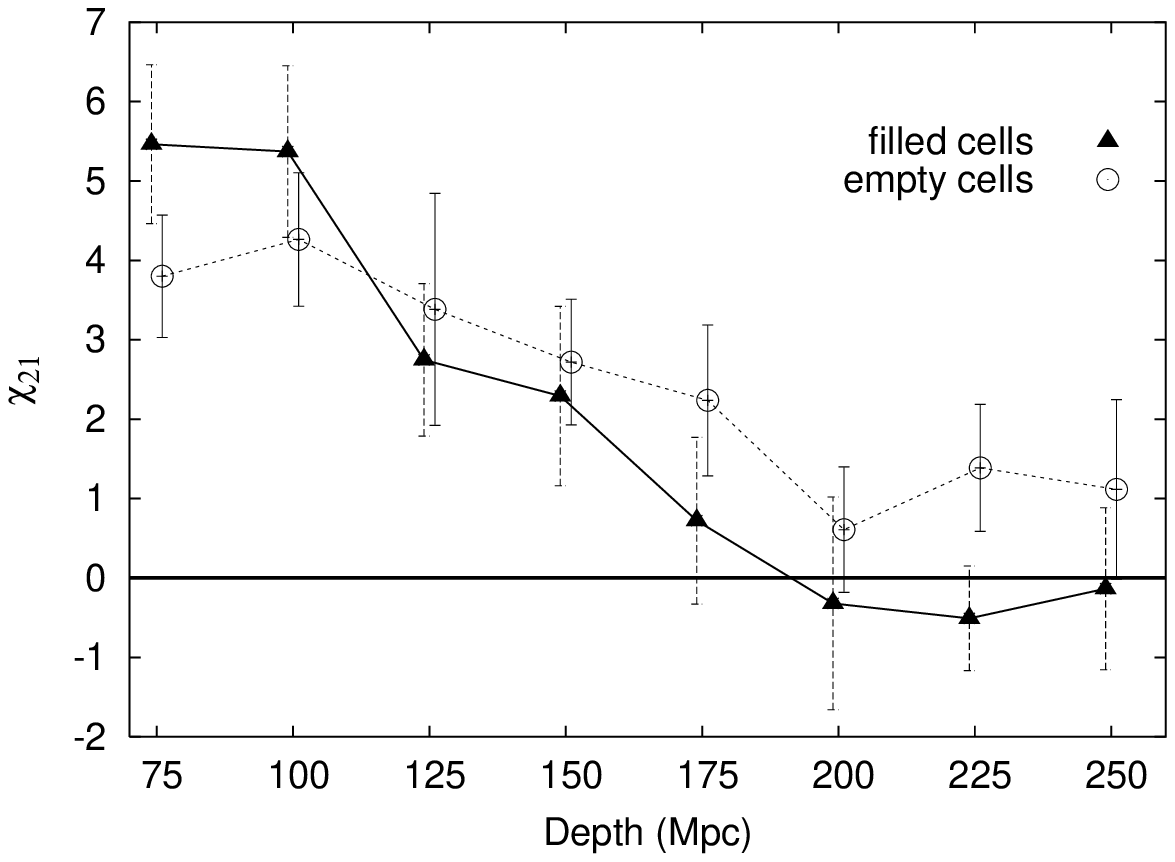}}
  \bc{\small Fig.\,6~~Mean $\chi_{21}$ as a function of the sample
depth. Same symbols as in Fig.\,3}\ec
\end{figure}

The variation of mean $\chi_{21}$ with the sample depth is
shown in Fig.\,6. For the filled cells, we see that there is
a considerable excess of ``sheet cells'' over ``string
cells'' at small sample depths. As the sample depth
increases, this excess becomes less and less, and eventually
vanishes altogether at around 200 Mpc. For the empty cells, there
is also a definite, though smaller, excess of ``sheet cells''
over ``string cells'' at 
the start, but the
decrease with increasing depth is gentler in this case, such
that even at the largest sample depths, there is still a
small residual excess.  Except for a short stretch at the
beginning, the empty curve lies consistently above the filled
curve, reminiscent of the $\av{\eta}$- curves of Fig.\,3.

\vn\vn{\bf 5.2~~Interpretation of Results}

So far we have described the trends of the indices as
variations with the sample depth, or the limiting distance of
the sample. But this is for verbal convenience only. We have
throughout this paper emphasized that each of our
distance-limited samples is also a luminosity-limited sample,
that as we consider deeper and deeper samples we are also
considering samples of more and more luminous galaxies.
Indeed, for the X-axis of our last four figures, we could equally
have 
used the limiting luminosity instead of the limiting
distance. So we must now address the question, ``which of the
two, limiting distance or limiting luminosity, is the 
the parameter the morphology correlates with ?''

This question is easily settled. All we have to do is to
consider the {\it intersection\/} of two of our samples, and
see how {\it its\/} results compare with those of the two 
original 
samples. We chose the intersection of S-100 and S-200. We
call it S$^{\ast}$-100, that is, the sample S$^{\ast}$-100 has
the same limiting distance as S-100 (100 Mpc) and the same
limiting luminosity as S-200 ($\log L_{\ast}=4.38$).

\bc {\small{\bf Table 4}~~Comparison of the Results of Three
Samples}\\
\fns
\begin{tabular}{ccc|cccc}

Sample & $\log L_{\ast}$ & f/e &
$\av{\eta}$ & $\av{\chi_1}$ & $\av{\chi_2}$ & $\av{\chi_{21}}$ \\ 
\hline\hline

S- 100 & 3.78 & f & 2.83 &-3.97 & 3.34 & 5.37\\
S$^{\ast}$-100 & 4.38 & f & 0.02 &-0.33 & 0.16 & 0.35\\
S- 200 & 4.38 & f &-0.40 & 0.29 &-0.17 &-0.32\\ \hline

S- 100 & 3.78 & e & 3.51 &-3.54 & 2.17 & 4.26\\
S$^{\ast}$-100 & 4.38 & e & 0.33 &-0.44 & 0.85 & 0.83\\
S- 200 & 4.38 & e & 0.77 &-0.38 & 0.46 & 0.61\\  \hline

\end{tabular}
\normalsize
\ec

Table 4 compares the mean indices from S$^{\ast}$-100 with those from
S-100 and S-200 lifted from Table 3, 
separately for the filled regions
(upper three lines) and 
the empty regions (lower three
lines). It is clear that the results of S$^{\ast}$-100 are much
closer to those of S-200 than to those of S-100: the
parameter that the morphology correlates with 
is {\it not\/} the sample depth; it is the limiting luminosity.

We must now re-state the empirical results obtained in the
present study in the following terms.  The morphology of the
filled and empty regions defined by a given
distance-and-luminosity-limited sample of galaxies is
essentially a function of the limiting luminosity.  As we
consider samples of more and more luminous galaxies at ever
decreasing 
space densities, the morphology
degrades: the degree of like cells flocking together steadily
decreases and so does the excess of cells belonging to thin
sheets over those belonging to thin strings. And the
degrading is generally stronger for the filled than for the
empty region, so that one could say, at any luminosity level,
the universe is always more like loose collections of lakes
in a land than groups of islands and archipelagos in an
ocean.

Recall that our luminosity is based on infrared flux and that its
definition at (3) is purely nominal, it may be useful for
future 
comparisons with results from other datasets
to introduce a more objective parameter than the limiting
luminosity $L_{\ast}$. We propose the space number density of
galaxies more luminous than $L_{\ast}$, to be denoted by
$\rho(L_{\ast})$. This quantity is easily calculated: for it
is simply the number of galaxies in the ``filled-out PSCz
Catalogue'' (Section 2) with $r<r_{\ast}$ and $L>L_{\ast}$,
divided by $(4/3)\pi r_{\ast}^{3}$.  The results (in galaxies
per (100\,Mpc)\up{3}) are given in the last column of Table
1.

\vn\vn{\bf 5.3~~A Density Effect or a Random Selection Effect
?}

It was suggested to us by the Referee that we should look
into the possibility that the observed variation in the
indices across the samples be a density effect (since, e.g.,
S-200 has a smaller density than S-100) and that, for testing
this possibility,  we should use a random sample of S-100, labelled
S$^{\ast\ast}$-100, with the same number of galaxies as S$^{\ast}$-100.  

As might be expected, the resulting indices varied much from
one random selection to the next. So we took 10 such random
selections. From now on, for simplicity, we shall restrict
the discussion to one of the indices, and we choose the
flocking index $\eta$ for the filled region. We found, for
the 10 random selections, $\eta$ ranges from -0.85 to +0.45,
with mean -0.24 and standard deviation 0.33. Let us denote
this mean value by $\eta^{\ast\ast}$; we have $\eta^{\ast\ast}=-0.24$. This
value is significantly smaller than the mean $\eta$ for
S-100, (which we now simply write as $\eta_1$ (from Table 3,
$\eta_1=2.63$)

Before interpreting this result ($\eta^{\ast\ast} \ll \eta_1$), we
should recall the reasoning behind our interpretation of a
previous, formally similar inequality, we mean the inequality
$\eta^{\ast} \ll \eta_1$, where $\eta^{\ast}$ stands for the mean
$\eta$ for the sample S$^{\ast}$-100 (according to Table 3,
$\eta^{\ast}=0.02$).  And we interpreted this latter inequality as
a luminosity effect because S*-100 simply consists of the
more luminous members of S-100.  Now, S$^{\ast\ast}$-100 is generated
out of S-100 quite differently: we just pick out a certain
prescribed number of its members, {\it purely at random\/},
without any regard to any individual properties. Hence, just
as we interpreted the inequality $\eta^{\ast} \ll \eta_1$ as a
luminosity effect, we must now interpret the inequality
$\eta^{\ast\ast} \ll \eta_1$ as an effect of random selection.

In fact, it seems quite plausible that any random selection
of a population will be less ``structured'' than the
population itself. Let us quantify the ``degree of 
structuredness'' by the ratio $F_2/F_1$, with $F_1$ the fraction
of isolated members and $F_2$ the fraction of members belonging
to groups of 2 or more. Then it seems obvious that if we make
a random selection of the population, with each member,
whether isolated or belonging to a group, having the same
chance of being selected, then the resulting $F_2/F_1$ will be
smaller. This point that random selection destroys structure
does not seem to have been noticed before in the literature;
it provides a natural interpretation of our present finding,
$\eta^{\ast\ast} \ll \eta_1$. 

\vspace{1cm} 
\no {\bf Brief Summary}~~The present study carries out
one of the programs outlined at the end of Paper I, the
application of rhombic cell analysis to a large size data.
But more has been done. Statistical indices, particularly a
``flocking index'', have been introduced to summarize much of
the essential information and, after emphasising the discrete
nature of the analysis and of the space distribution of
galaxies, these indices for any given sample of galaxies are
shown each to have an irreducible scatter. Despite the
scatter, the indices showed remarkable variation with the
limiting luminosity of the sample, leading to the conclusion
stated above. It is planned to carry out further programs
mentioned in Paper I, particularly the raising of the
threshold of the filled cell.

\vspace{1cm}
\no{\bf Acknowledgements}~~We thank the Referee for making
a suggestion which led to a new insight as reported in
Subsection 5.3. We thank Dr F. Fontanot of Trieste for making
a list of mock galaxies available to us.

\vspace{1cm}

\vn\vn {\bf REFERENCES}

\no Branchini E., Teodoro L., Frenk C. S., et al., 1999, MNRAS, 308, 1

\no Fontanot F., Monaco P., Borgani S., 2003, MNRAS, 341, 692

\no Kiang T., 2003, Chin.\ J.\ Astron.\ Astrophys., 3/2, 95-104

\no Saunders W., Sutherland W. J., et al., 2000, MNRAS, 317,
55-63

\end{document}